# Non-Hermitian stealthy hyperuniformity

Gitae Lee[1‡], Seungmok Youn[1‡], Ikbeom Lee[1‡], Kunwoo Park[1], Duhwan Hwang[1], Xianji Piao[2§], Namkyoo Park[3†], and Sunkyu Yu[1*]

[1]Intelligent Wave Systems Laboratory, Department of Electrical and Computer Engineering, Seoul National University, Seoul 08826, Korea

[2]Wave Engineering Laboratory, School of Electrical and Computer Engineering, University of Seoul, Seoul 02504, Korea

[3]Photonic Systems Laboratory, Department of Electrical and Computer Engineering, Seoul National University, Seoul 08826, Korea

E-mail address for correspondence: §piao@uos.ac.kr, †nkpark@snu.ac.kr, *sunkyu.yu@snu.ac.kr

**Abstract**

Symmetry-driven wave physics in open systems, exemplified by parity-time (PT) symmetry, has extended the landscape of crystalline phases in materials science to include gain-loss media. Given the growing interest in engineering disorder for wave manipulation, such non-Hermitian crystals motivate the extension of non-Hermitian frameworks into the realm of correlated disorder. Here, we propose hyperuniformity and stealthiness in non-Hermitian systems as a generalization of PT-symmetric crystals to correlated disorder. We extend the scattering-microstructure correspondence to open systems, formulating non-Hermitian hyperuniformity and stealthiness that encompass their Hermitian counterparts. This approach—incorporating a statistical crystallography framework for non-Hermitian materials—demonstrates that real-imaginary cross-correlations of the material potential are irrelevant for achieving hyperuniformity but are essential for characterizing stealthiness, revealing unidirectional scattering phases that are inaccessible in Hermitian materials and in non-Hermitian crystals. By analysing the microstructural statistics of the resulting materials, our results—building on non-Hermitian wave physics—establish a connection to materials science, encompassing conventional descriptors of correlated disorder.

# Introduction

Bridging non-Hermitian wave physics[1,2]—exemplified by parity-time (PT) symmetry[3]—with discrete translational symmetry has rejuvenated crystallography by extending traditional band theory to systems composed of periodically arranged gain and loss media. This generalized band theory is founded on concepts underlying open-system wave phenomena: biorthogonal bulk-boundary correspondence[4], nontrivial topology around exceptional points[5,6], and generalized Brillouin zones[7]. Building on these well-established analytical tools, the traditional realm of crystals has been extended into non-Hermitian regimes, unveiling exotic wave phenomena in PT-symmetric crystals, such as non-Hermitian skin effects[8], unidirectional transparency[9] and modal conversion[10], and non-Abelian band braiding[11].

In classifying material phases according to their microstructures, the regime of order has been substantially extended through the concepts of hyperuniformity (HU)[12,13] and stealthiness[14]. Originally developed in microstructural statistics to describe the vanishing of long-wavelength density fluctuations[13,15], these concepts exhibit an intriguing correspondence with wave phenomena[16]—the suppression of long-wavelength scattering over a range of reciprocal space in the weak-scattering regime. This correspondence has revealed numerous previously unrecognized wave phenomena, including perfect isotropic bandgaps[17-20] and transparency[21], symmetry-free guiding and resonances[22,23], the screening of material microstructures[24], and patternless spectral filtering[25]. Given the extension from traditional Hermitian crystals to PT-symmetric crystals from a wave-physics perspective, and the interpretation of HU and stealthiness as statistically generalized order, a natural question arises: what is the non-Hermitian generalization of HU and stealthiness? To address this question, we can envisage the use of the correspondence between microstructural statistics and scattering phenomena in open systems, in line with previous studies of Hermitian systems.

Here, we propose non-Hermitian generalizations of hyperuniformity (HU) and stealthiness by leveraging the scattering-microstructure correspondence. By characterizing the conditions for suppressed scattering in non-Hermitian materials, we define HU and stealthiness in the presence of non-Hermiticity. This approach reveals that cross-correlations between the real and imaginary parts of the material potential are critical only for stealthiness. We develop a statistical crystallography framework to classify non-Hermitian stealthiness by exploiting the rotational symmetries of these correlations, thereby unveiling novel material phases with unidirectional scattering responses. To connect these wave-physics-based concepts to established material phases, we also examine the microstructural statistics of non-Hermitian HU and stealthy HU (SHU). Our results, extending correlated disorder to open systems, provide new design freedom for wave functionality, particularly through non-Hermiticity-induced directionality.

## Results

**Non-Hermitian HU**

In Hermitian systems, a HU material with suppressed long-wavelength density fluctuations corresponds to scattering suppression near zero momentum shift, $\mathbf{k} \cong 0$, under the first-order Born approximation[12,13,16]. Motivated by this correspondence, we extend HU to non-Hermitian wave physics by analysing the conditions for scattering suppression in a nonconservative composite material containing gain and loss media (Fig. 1a). We begin with the wave equation $\nabla^2 \psi + V(\mathbf{r})\psi = 0$ for a complex-valued potential $V(\mathbf{r})$ that is spatially confined to a finite domain $\Omega$: $V(\mathbf{r} \in \Omega) = V_o + V_a(\mathbf{r})$ and $V(\mathbf{r} \in \Omega^c) = V_o$ with a constant, real-valued bias potential, $V_o > 0$. For the planewave incidence $\psi_{in}(\mathbf{r}) = \psi_I \exp(i\mathbf{k}_I \cdot \mathbf{r})$ with $|\mathbf{k}_I|^2 = V_o$, the Lippmann-Schwinger equation becomes

$$\psi(\mathbf{R}) = \psi_{in}(\mathbf{R}) + \frac{1}{4\pi}\int G(\mathbf{R},\mathbf{r};\mathbf{k}_S)V_a(\mathbf{r})\psi(\mathbf{r})d\mathbf{r},$$

where $G(\mathbf{R},\mathbf{r};\mathbf{k}_S)$ is the Green's function and $\mathbf{k}_S$ denotes the scattering wavevector[26]. Under the Born approximation, the far-field scattering intensity is well approximated as $I_S(\mathbf{R}) \propto S(\mathbf{k} \triangleq \mathbf{k}_S - \mathbf{k}_I)|\psi_I|^2$ using the structure factor $S(\mathbf{k}) = S_A(\mathbf{k}) + S_C(\mathbf{k})$, defined as follows (Supplementary Note S1):

$$S_A(\mathbf{k}) = \frac{1}{V_\Omega}\left[\langle|V_r(\mathbf{k})|^2\rangle + \langle|V_i(\mathbf{k})|^2\rangle\right], \quad S_C(\mathbf{k}) = \frac{2}{V_\Omega}\text{Im}\left[\langle V_r(\mathbf{k})V_i^*(\mathbf{k})\rangle\right], \tag{1}$$

where $V_r(\mathbf{k})$ and $V_i(\mathbf{k})$ are the reciprocal-space potentials of the real and imaginary parts of $V_a(\mathbf{r})$, respectively, $V_\Omega$ is the volume of $\Omega$, and $\langle\ldots\rangle$ denotes the statistical average under the ergodic condition[27]. Notably, nonconservative perturbations with $\text{Im}[V_a(\mathbf{r})] \neq 0$ contribute to scattering through the auto- and cross-correlations, $\langle|V_i(\mathbf{k})|^2\rangle$ and $\langle V_r(\mathbf{k})V_i^*(\mathbf{k})\rangle$, respectively.

We emphasize that scattering from a non-Hermitian material is constrained by the conjugate symmetries of the complex-valued potentials in reciprocal space: $V_r(-\mathbf{k}) = V_r^*(\mathbf{k})$ and $V_i(-\mathbf{k}) = V_i^*(\mathbf{k})$. Applying these symmetries to Eq. (1) yields contrasting parity conditions under inversion: $S_A(-\mathbf{k}) = S_A(\mathbf{k})$ and $S_C(-\mathbf{k}) = -S_C(\mathbf{k})$, which is analogous to the PT-symmetric condition of the real and imaginary parts of the potential[1-3]. Applying the scattering-suppression criterion for HU, the non-Hermitian generalization of HU requires the following condition:

$$\lim_{|\mathbf{k}|\to 0}\left[S_A(\mathbf{k}) + S_C(\mathbf{k})\right] = \lim_{|\mathbf{k}|\to 0} S_A(\mathbf{k}) = 0, \tag{2}$$

because $|S_A(\mathbf{k})| \geq |S_C(\mathbf{k})|$ for all $\mathbf{k}$. Therefore, Eqs. (1) and (2) leads to the definition of non-Hermitian HU (Fig. 1b)—simultaneous HU of the real and imaginary parts of the potential $V_a(\mathbf{r})$, such that $|V_r(\mathbf{k})|^2 \to 0$ and $|V_i(\mathbf{k})|^2 \to 0$ as $|\mathbf{k}| \to 0$, remarkably, regardless of their relative spatial distributions.

To elucidate this definition of non-Hermitian HU, we present a variety of point-particle material microstructures in Figs. 1c-1e (see 'Non-Hermitian scatterers' in Methods) together with their structure-factor profiles (Figs. 1f-1k), which comprise real-valued scatterers ($\text{Im}[V_a(\mathbf{r})] = 0$) and imaginary-valued gain-loss scatterers ($\text{Re}[V_a(\mathbf{r})] = 0$). When HU features are imposed on both the real and imaginary particles (Fig. 1c,d), long-wavelength scattering is successfully suppressed with $I_S(\mathbf{R}) \cong S(\mathbf{k}\to 0) \cong 0$ (Fig. 1f,g,i,j), in sharp contrast to the case where a non-HU configuration present in either the real or imaginary particle profile (Fig. 1e,h,k). This observation demonstrates that non-Hermitian systems provide extended degrees of freedom for altering the scattering features (Fig. 1f,g) via distinct microstructures of the real and imaginary HU potentials and their cross-correlation (Figs. 1c,d), while strictly preserving HU suppression of long-wavelength scattering.

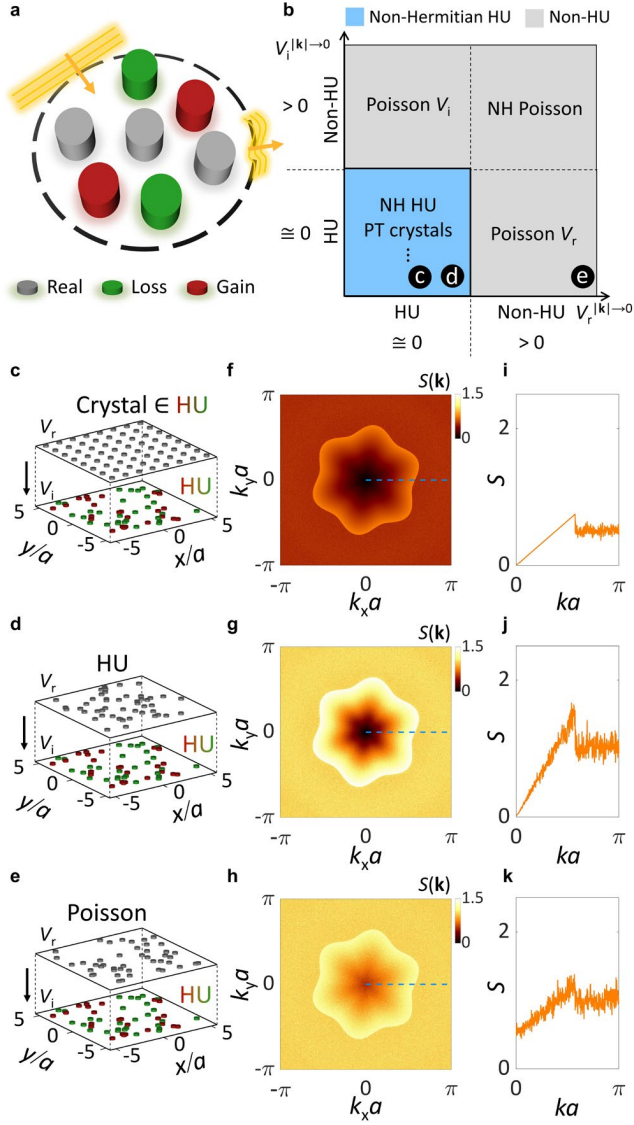

**Fig. 1 Non-Hermitian HU**. **a**, Schematic illustrating scattering from a multiparticle non-Hermitian material. Grey, red, and green rods represent real-valued, purely gain, and purely loss scatterers, respectively. **b**, Phase diagram characterizing non-Hermitian HU. The phases are classified with $V_r(|\mathbf{k}| \to 0)$ and $V_i(|\mathbf{k}| \to 0)$; 'NH HU' denotes non-Hermitian HU; 'PT' represents PT symmetry; 'Poisson $V_r$' and 'Poisson $V_i$' represent real and imaginary Poisson potentials, while preserving $V_i(|\mathbf{k}| \to 0) \cong 0$ and $V_r(|\mathbf{k}| \to 0) \cong 0$, respectively; 'NH Poisson' denotes the Poisson distributions for both $V_r$ and $V_i$. **c-e**, Examples of $N$-particle non-Hermitian materials: 'NH HU' materials (**c,d**) and 'Poisson $V_r$' (**e**). **f-k**, The corresponding structure factors (**f-h**) and their cross-sections (**i-k**; blue dashed lines in **f-h**). $a = L/N^{1/2}$ for a square supercell of side length $L$ in **c-k**.

**Density fluctuations of non-Hermitian HU**

To confirm the validity of the proposed non-Hermitian HU, the resulting material microstructures should allow a rational interpretation from a materials-science perspective, especially encompassing the traditional definition of HU in Hermitian systems. We therefore examine density fluctuations in non-Hermitian materials across different material phases, focusing on the uniqueness of non-Hermitian HU. In Hermitian systems, HU microstructural statistics manifest as suppressed number-density fluctuations, such as $\sigma^2(R)/R^d \sim 1/R$, in sharp contrast to the constant $\sigma^2(R)/R^d$ observed in uncorrelated disorder[13], where $R$ is the radius of the sampling window used to measure number-density fluctuations and $d$ denotes the system dimensionality. For a complex-valued potential $V_a(\mathbf{r})$, this density-based analysis must be generalized into three quantities (see 'Density fluctuations' in Methods): the auto-density fluctuations, $\sigma_{AR}^2(R)/R^d$ and $\sigma_{AI}^2(R)/R^d$, which denote the volume-normalized variances of $\text{Re}[V_a(\mathbf{r})]$ and $\text{Im}[V_a(\mathbf{r})]$, respectively, and the cross-density fluctuation, $\sigma_C^2(R)/R^d$, which is the volume-normalized covariance between $\text{Re}[V_a(\mathbf{r})]$ and $\text{Im}[V_a(\mathbf{r})]$.

Figure 2 examines our wave-physics-based definition of non-Hermitian HU through a microstructural analysis (see 'Density measurements' in Methods; Supplementary Note S2 for semi-analytical demonstration). For each realization, we sample the potential $V_a(\mathbf{r})$ of an $N$-particle material using a circular window of radius $R$ to compute the normalized density-fluctuation measures, $\sigma_{AR}^2(R)/(\rho A)$, $\sigma_{AI}^2(R)/(\rho A)$, and $\sigma_C^2(R)/(\rho A)$, where $\rho$ is the number density of scatterers and $A(R) = \pi R^2$ is the window area (see Supplementary Note S2 for analytical expressions in terms of pair correlation functions). We compare different types of non-Hermitian HU materials—a PT-symmetric crystal (Fig. 2a) and a disordered non-Hermitian HU material (Fig. 2b)—with a non-Hermitian Poisson material defined in Fig. 1b (Fig. 2c).

The results demonstrate that both the PT-symmetric crystal and the disordered non-Hermitian HU material exhibit suppressed auto-density fluctuations as $R$ increases (Fig. 2a,b), which underlies the long-wavelength scattering suppression $S(|\mathbf{k}|\to 0) = 0$. In contrast, the Poisson configuration exhibits $R$-independent plateaus in the auto-density fluctuations (Fig. 2c), a hallmark of uncorrelated disorder. These results confirm that non-Hermitian HU preserves the conventional scaling behaviour of density fluctuations for the real and imaginary parts of the potential separately. Notably, the near-zero cross-density fluctuations observed in our configurations originate from the balanced numbers of gain and loss particles and their matched statistical distributions, remaining further design freedom to alter microstructures while preserving HU (Supplementary Note S2).

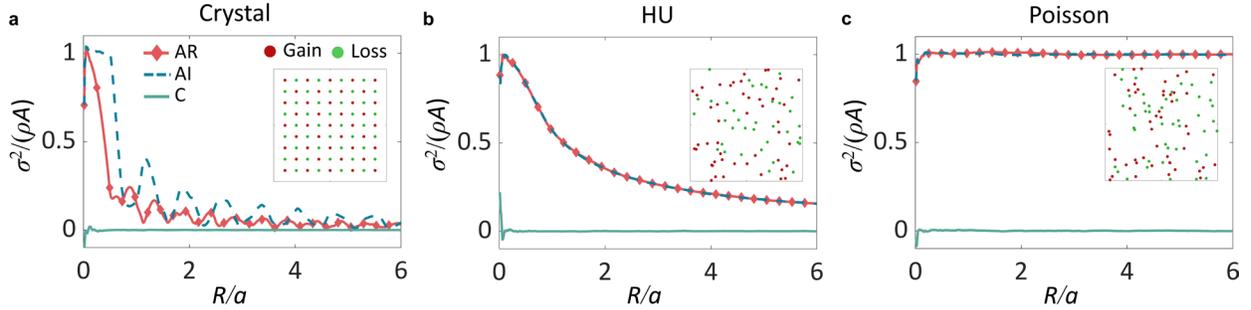

**Fig. 2. Density fluctuations in non-Hermitian materials**. **a-c,** Density-fluctuations in a PT-symmetric square-lattice crystal (**a**), a disordered non-Hermitian HU (**b**), and an uncorrelated non-Hermitian Poisson material (**c**). Each inset shows the portion of representative material realizations. Coordinates and window radius are normalized by $a = L/N^{1/2}$. Variances are calculated over an ensemble of 100 realizations. The cinnabar-marker and blue-dashed curves denote the volume-normalized variances of the real and imaginary potentials, respectively, while the teal-solid curve represents the volume-normalized covariances between them.

**Non-Hermitian SHU**

Beyond the non-Hermitian generalization of HU, we explore a non-Hermitian generalization of stealthiness[14], which is defined by $S(|\mathbf{k}| \leq K) \approx 0$ in Hermitian systems for a finite positive scalar $K$. We note that suppression of scattering over $|\mathbf{k}| \leq K$ is governed by inversion symmetries, $S_A(-\mathbf{k}) = S_A(\mathbf{k})$ and $S_C(-\mathbf{k}) = -S_C(\mathbf{k})$. In contrast to non-Hermitian HU, these symmetries impose angular constraints on realizing non-Hermitian SHU. To characterize these constraints, we develop a statistical crystallography framework for classifying non-Hermitian correlated disorder by examining the rotational symmetries of statistical metrics relevant to long-wavelength scattering.

In our framework, we examine the rotational symmetries encoded in the angular structure of the constraints $S_A(|\mathbf{k}| \leq K) \approx 0$ and $S_C(|\mathbf{k}| \leq K) \approx 0$ (Fig. 3a)—including the symmetries of the Bragg peaks at $|\mathbf{k}| \approx K$ in crystalline structures. Conventional Hermitian materials (boxes shaded in teal in Fig. 3a), which always satisfy $S_C(\mathbf{k}) = 0$, span a parameter range in which the allowed symmetry classes of $S_A(\mathbf{k})$ are $C_{2m}$, while those of $S_C(\mathbf{k})$ are $C_\infty$, where $C_n$ denotes the group of $n$-fold rotational symmetry. Although square, triangular, and honeycomb lattices—representing the 2D crystals accessible via regular polygon tiling—occupy only a subset of this range (yellow stars in Fig. 3a), SHU spans a broader set of symmetry classes, including all $C_{2m}$ symmetries attainable with isotropic[13] and anisotropic SHU[28] materials.

In contrast, we emphasize that non-Hermitian configurations with a nonzero cross-correlation $S_C(\mathbf{k})$—which exhibits inherent $C_{2m+1}$ symmetries—substantially extend the accessible range of scattering responses (boxes shaded in cinnabar in Fig. 3a). In Figs. 3b-3g, we illustrate the structure factors $S(\mathbf{k})$ of a variety of non-Hermitian ordered and disordered structures, including disordered SHU phases inversely designed using ansatz functions[29,30] (Figs. 3b-3d; see 'Non-Hermitian SHU ansatz' and 'Inverse design' in Methods and Supplementary Note S3) and examples of PT-symmetric crystals (Figs. 3e-3g; Supplementary Notes S4 and S5 for extended analysis with full-wave simulations[31]). Notably, non-Hermitian materials can exhibit long-wavelength scattering phases identical to those obtained with Hermitian systems: conventional

SHU (Fig. 3b), anisotropic SHU (Fig. 3c)[28], a $C_4$-crystal (Fig. 3e), and an anisotropy-induced $C_2$-crystal (Fig. 3f). Such isoscattering responses in the region $|\mathbf{k}| \leq K$ provide design freedom to tailor other wave characteristics, such as scattering at $|\mathbf{k}| > K$ or the spectral bandwidth[16,24].

More importantly, non-Hermitian degrees of freedom enable rotational-symmetry classes in long-wavelength scattering that are inaccessible in Hermitian systems—for example, $C_3$-crystals (Fig. 3g), and $C_3$ disordered SHU materials (Fig. 3d), which satisfy $S(\mathbf{k}) \neq S(-\mathbf{k})$ and can even support unidirectional scattering. This result—while remaining consistent with electromagnetic reciprocity in the absence of time-reversal symmetry breaking[32]—identifies the microstructural conditions in non-Hermitian correlated disorder that lead to asymmetric scattering, a phenomenon that has been intensively studied in PT-symmetric crystals[1].

We note that the angular symmetries of the scattering observed in Fig. 3 highlight the necessity of interpreting the SHU condition for non-Hermitian materials in terms of angularly restricted, lower-dimensional configurations. Remarkably, because $S_C(\mathbf{k})$ enlarges the space of possible angular responses in the overall scattering, it is useful to consider a 1D SHU condition of the form of $S(|\mathbf{k}| \leq K; \theta(\mathbf{k}) \in \Phi) \approx 0$ in our 2D example, where $\theta(\mathbf{k})$ is the polar angle of $\mathbf{k}$ and $\Phi$ denotes the set of angles over which the SHU condition is satisfied. In this formulation, $\Phi$ is determined by the $C_n$ symmetries of $S_A(\mathbf{k})$ and $S_C(\mathbf{k})$, whereas the conventional 2D SHU—corresponding to the full angular range of $\Phi = [0, 2\pi)$—is recovered when both $S_A(\mathbf{k})$ and $S_C(\mathbf{k})$ belong to $C_\infty$.

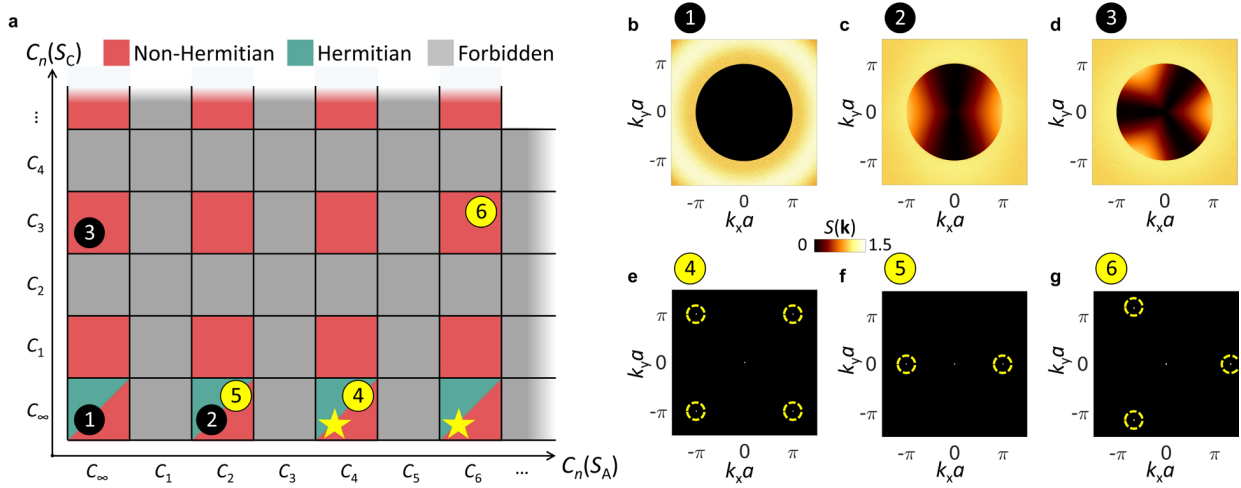

**Fig. 3. Statistical crystallography for non-Hermitian order and SHU. a**, Classification of scattering responses by the rotational symmetries of $S_A(\mathbf{k})$ and $S_C(\mathbf{k})$. Cinnabar and teal colours denote scattering responses that can be realized in non-Hermitian and Hermitian materials, respectively. Yellow stars denote 2D Hermitian crystals with regular polygon tiling. 'Forbidden' indicates inaccessible scattering responses due to inherent constraints on $S_A(\mathbf{k})$ and $S_C(\mathbf{k})$. **b-d**, Non-Hermitian SHU with $C_\infty$ (**b**), $C_2$ (**c**), and $C_3$ (**d**) symmetries. **e-g**, PT-symmetric crystals with $C_4$ (**e**), $C_2$ (**f**), and $C_3$ (**g**) symmetries. All materials consist of $10^6$ scatterers. The structure factor is obtained through the average of $10^2$ ensemble realizations.

**Density fluctuations of non-Hermitian SHU**

Along with the analysis in Fig. 2, we examine the material microstructures of the obtained non-Hermitian SHU, thereby connecting our definition to a material-science interpretation. To capture angularly varying density fluctuations in non-Hermitian multiphase materials, we generalize the conventional number-variance measure by using two correlated observation windows 1 and 2 with the same radius $a$ (Fig. 4a, Supplementary Note S6), which is sufficiently small to probe local distributions. The displacement from the centre of window 1 to that of window 2 is expressed in polar coordinates $(r, \varphi)$. Using the windows, we measure the integrated amount of real and imaginary parts of the potential perturbation within window $n$ (= 1, 2), denoted as $N_{n,\mathrm{r}}$ and $N_{n,\mathrm{i}}$, respectively. Pairs of windows are placed uniformly at random throughout the unit cell of the material for $10^5$ trials, to obtain stable statistics of ($N_{1,r}$, $N_{1,i}$, $N_{2,r}$, $N_{2,i}$).

Figures 4b-4g show the microstructural statistics of non-Hermitian disordered SHUs in Fig. 3b-3d, for their auto- ($\Sigma_A^2 = \langle N_{1,r}N_{2,r}\rangle + \langle N_{1,i}N_{2,i}\rangle$; Fig. 4b,d,f) and cross- ($\Sigma_C^2 = \langle N_{1,i}N_{2,r}\rangle - \langle N_{1,r}N_{2,i}\rangle$; Fig. 4c,e,g) covariances of real and imaginary potentials. Notably, the structure factors accessible in Hermitian systems (Fig. 3b,3c) correspond to the complete suppression of real-imaginary covariances $\Sigma_C^2$ (Fig. 4c,e), as expected from the vanishing scattering interference with $S_C(\mathbf{k})$. On the other hand, the unidirectional scattering response in Fig. 3d corresponds to the emergence of $\Sigma_C^2$ (Fig. 4g). We note that the emergence of the characteristic length $\sim 2a$ and its antisymmetric profile of $\Sigma_C^2$ indicate that our non-Hermitian disordered SHUs correspond to a disordered composite of PT-symmetric dipoles.

The proposed approach for capturing microstructural features of non-Hermitian materials reduces to a coarse-grained representation of the conventional two-point correlation function in the limit of a vanishing window size. However, the method differs conceptually and practically by probing finite-support correlations at a prescribed mesoscopic scale. By integrating local quantities over observation windows, the method is robust against pointwise noise, sharp interfaces, and phase discontinuities, and is well suited for both numerical simulations and experimental measurements of non-Hermitian materials. Importantly, explicit control of the window size enables the characterization of angularly varying features at a target mesoscopic length scale, complementing the information obtained from pointwise correlation analysis.

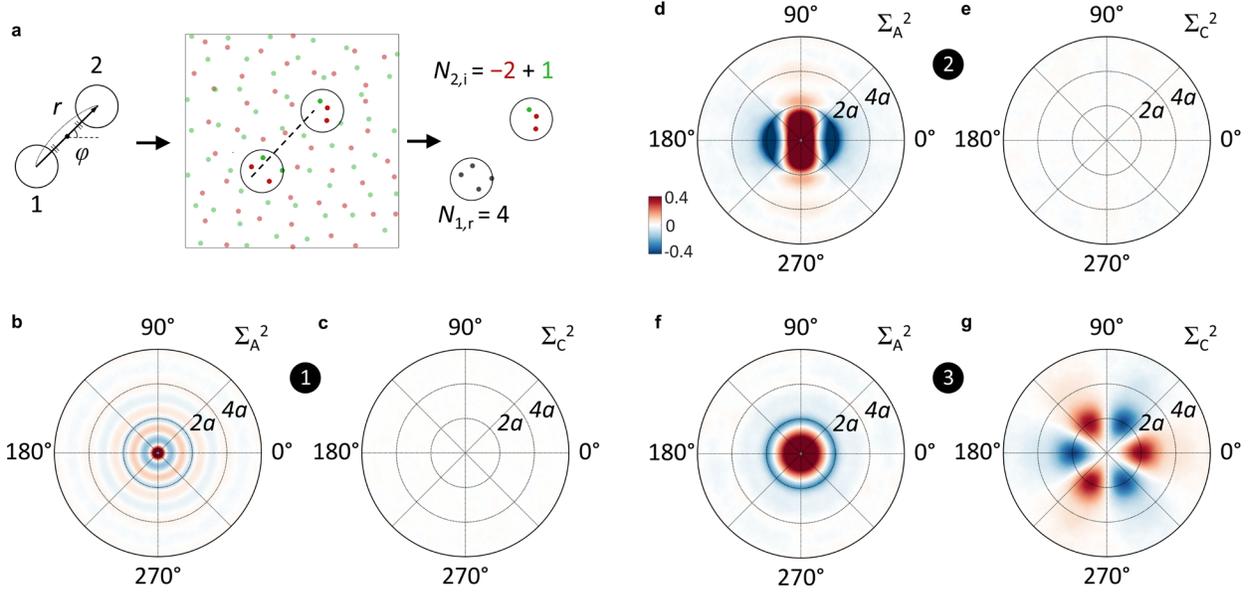

**Fig. 4. Directional number statistics. a,** Extraction of directional number statistics using two circular, correlated windows with fixed radii of characteristic length *a*. The displacement between two windows is swept with parameters $r$ and $\varphi$. Window pairs are placed uniformly at random in the unit cell for $10^5$ trials for each ($r$, $\varphi$). **b-g,** Auto- and cross-covariances of non-Hermitian SHU under $C_\infty$ (**b,c**), $C_2$ (**d,e**), and $C_3$ (**f,g**) symmetries described in Fig. 3. Radial distance and azimuthal angle of polar plots correspond to $r$ and $\varphi$, respectively.

## Discussion

Because our disordered non-Hermitian HU systems can be viewed as statistical generalizations of PT-symmetric crystals, the results suggest a route toward extending non-Hermitian band theory into the regime of correlated disorder. Analogous to the extension of topological physics to HU materials[33], we can map the point-process realization of gain and loss scatterers onto the positions of discretized elements in a tight-binding description. With this approach, we can extend phenomena arising from non-Hermitian band theory to non-Hermitian SHU systems. For example, the anisotropy of exceptional rings that is unavoidable in non-Hermitian crystals[34] may be mitigated—and potentially replaced by perfectly isotropic rings—in non-Hermitian SHU. Further studies in directions related to topological Anderson insulators[35] may also be of interest using correlated features in SHU.

In terms of a practical implementation, our study relies on the Born approximation in the weak-scattering regime, which is well suited to dilute or low-contrast non-Hermitian media. Considering the challenge in handling gain scatterers, one may envisage passive demonstrations based on a gauge transformation[36] or the use of an amplifying environment including subwavelength lossy scatterers[37]. In this context, further studies in the multiple scattering regime are also desired, particularly regarding wave localization in non-Hermitian media[38]. By conceptualizing gain and loss scatterers as distinct material phases, our work also extends the framework of SHU in multiphase media incorporating open-system design freedom, which holds potential for multiphysics applications[39-41].

In summary, we generalize HU and stealthiness to non-Hermitian wave systems by establishing a wave-based definition grounded in scattering suppression and microstructural statistics. Our framework shows that non-Hermitian HU requires independent suppression of long-wavelength fluctuations in both the real and imaginary parts of the potential, while non-Hermitian SHU depends crucially on their cross-correlations. We develop the rotation-symmetry-based crystallography for non-Hermitian systems, enabling the characterization of unidirectional scattering responses in disordered non-Hermitian media. By bridging non-Hermitian wave physics and material microstructures, our work extends PT-symmetric concepts beyond periodic lattices in scattering phenomena, paving the way toward directional wave manipulations in optics, acoustics, and electronic circuits.

## Methods

**Non-Hermitian scatterers.** In our analysis, we employ the point-particle approximation, which yields the structure factor for an $N$-particle material:

$$S(\mathbf{k}) = \frac{1}{\sum_{p=1}^{N}|f_p|^2}|C(\mathbf{k})|^2, \tag{3}$$

where $C(\mathbf{k}) \triangleq \sum_p f_p \exp(-i\mathbf{k}\cdot\mathbf{r}_p)$ represents the collective coordinate variable, and $\mathbf{r}_p$ and $f_p = f_{r,p} + if_{i,p}$ denote the position and complex-valued scattering form factor of the $p$th scatterer, respectively, with $f_{r,p}, f_{i,p} \in \mathbb{R}$. For the results in Fig. 1, we set $f_p \in \{1, -i, +i\}$, while we use $f_p \in \{1-i, 1+i\}$ in Figs. 2-4 except for Fig. 3g with $f_p \in \{1 - i/3^{1/2}, 1 + i/3^{1/2}\}$ for maximized unidirectionality. Especially, we can express the form-factor components as $f_{r,p} = f_r$ and $f_{i,p} = s_p f_i$, by setting $s_p \in \{-1, +1\}$ in Figs. 2-4 and $f_i = 1/3^{1/2}$ in Fig. 3g. The applied condition ensures that the contributions of auto- and cross-correlations to scattering are comparable in magnitude.

**Density fluctuations.** For the analysis in Fig. 2, we introduce the indicator function $w(\mathbf{r} - \mathbf{x}_0; R)$ associated with a circular observation window of radius $R$ centred at $\mathbf{x}_0$. This function is defined as $w(\mathbf{r} - \mathbf{x}_0; R) = 1$ if a scatterer is located within the window and $w(\mathbf{r} - \mathbf{x}_0; R) = 0$ otherwise. Using this indicator function, the complex-valued potential perturbation from scatterers within the window is expressed as $N_c = N_r + iN_i$, where

$$N_r(\mathbf{x}_0; R) = \sum_p f_r w(\mathbf{r}_p - \mathbf{x}_0; R), \quad N_i(\mathbf{x}_0; R) = \sum_p s_p f_i w(\mathbf{r}_p - \mathbf{x}_0; R). \tag{4}$$

The normalized density fluctuation—defined as the variance of the potential perturbation normalized by the total perturbation—is generalized as

$$\frac{\langle N_c^2 \rangle - \langle N_c \rangle^2}{\rho A(R)} = \frac{\sigma_{AR}^2(R)}{\rho A(R)} - \frac{\sigma_{AI}^2(R)}{\rho A(R)} + 2i\frac{\sigma_C^2(R)}{\rho A(R)}, \tag{5}$$

for the variance and covariance expressions:

$$\sigma_{AR}^2(R) = \text{cov}[N_r(\mathbf{x}_0; R), N_r(\mathbf{x}_0; R)],$$
$$\sigma_{AI}^2(R) = \text{cov}[N_i(\mathbf{x}_0; R), N_i(\mathbf{x}_0; R)], \qquad (6)$$
$$\sigma_C^2(R) = \text{cov}[N_r(\mathbf{x}_0; R), N_i(\mathbf{x}_0; R)],$$

where cov[$a,b$] denotes the covariance between the random variables $a$ and $b$ over an ensemble of random window positions $\mathbf{x}_0$.

**Density measurements.** In our material configuration, point scatterers are distributed within a square supercell of side length $L$ under periodic boundary conditions, with the number density of $\rho = N/L^2$ and the characteristic length of $a = 1/\rho^{1/2}$. Auto- and cross-densities are estimated using the Monte Carlo window sampling[13]. For a given window centre $\mathbf{x}_0$ and radius $R$, we compute the sums of the potential perturbations inside the window using Eq. (4). Using the obtained $N_r$ and $N_i$, we numerically calculate the auto- and cross-densities by estimating their variances and covariance.

**Non-Hermitian SHU ansatz.** Consider a structure factor expressed as a superposition of functions possessing $m$-fold rotational symmetry:

$$S(\mathbf{k}) = \sum_{q \geq 0} \left[ A_q(|\mathbf{k}|) \cos(qm\theta(\mathbf{k})) + B_q(|\mathbf{k}|) \sin(qm\theta(\mathbf{k})) \right]. \qquad (7)$$

We retain the zeroth ($q = 0$) and first harmonic ($q = 1$) terms while imposing nonnegativity constraint $S(\mathbf{k}) \geq 0$ and assuming a common $|\mathbf{k}|^\alpha$ dependence for SHU[13], where $\alpha > 0$ is the power-law exponent characterizing the degree of SHU for $S(|\mathbf{k}| \leq K) \approx 0$. The result yields the following ansatz of $S_A(\mathbf{k})$ and $S_C(\mathbf{k})$:

$$S_A(\mathbf{k}) \sim \frac{1}{2}|\mathbf{k}|^\alpha \left( 1 + \frac{1}{2}\left[1 + (-1)^m\right]\cos(m\varphi_{\mathbf{k}}) \right),$$
$$S_C(\mathbf{k}) \sim \frac{1}{2}|\mathbf{k}|^\alpha \left( \frac{1}{2}\left[1 - (-1)^m\right]\cos(m\varphi_{\mathbf{k}}) \right). \qquad (8)$$

The resulting structure factor $S(|\mathbf{k}| \leq K) = |\mathbf{k}|^\alpha [1 + \cos(m\varphi_\mathbf{k})]/2$ provides an example of scattering responses with $m$-fold rotational symmetry, as described in Fig. 3.

**Inverse design.** Under periodic boundary conditions on a square supercell of side length $L$, the set of $\mathbf{k}$ contributing to the structure factor is restricted to a finite discrete set, $B = \{\mathbf{k} \mid \mathbf{k} = 2\pi(n_x, n_y)/L,\ n_x, n_y \in \mathbb{Z},\ 0 < |\mathbf{k}| \leq K\}$. To optimize point patterns with a target structure factor $S(\mathbf{k})$, we define the loss function using a weighted least-squares:

$$\mathcal{L}(\mathbf{r}_1, \ldots, \mathbf{r}_N) = \sum_{\mathbf{k} \in B} \frac{(S(\mathbf{k}) - S_0(\mathbf{k}))^2}{|\mathbf{k}|^2}, \qquad (9)$$

which assigns larger weights to longer-wavelength suppression. Its gradient is given by

$$\frac{\partial \mathcal{L}}{\partial \mathbf{r}_p} = \frac{4}{\sum_{j=1}^{N} |f_j|^2} \operatorname{Im}\left[ \sum_{\mathbf{k} \in B} \frac{\mathbf{k}}{|\mathbf{k}|^2} (S(\mathbf{k}) - S_0(\mathbf{k})) C^*(\mathbf{k}) f_p e^{-i\mathbf{k} \cdot \mathbf{r}_p} \right]. \qquad (10)$$

where $C^*(\mathbf{k})$ denotes the complex conjugate of the collective coordinate variable. We use the Flatiron Institute Nonuniform Fast Fourier Transform (FINUFFT) for the type-1 transform to compute collective coordinate variables and the type-2 transform to evaluate the gradient[29,30]. Each update direction is obtained using conjugate gradient method with a step size determined by a line search satisfying the Wolfe condition.

## Data availability

Data used in the current study are available from the corresponding authors upon request and can also be obtained by running the shared codes at 10.5281/zenodo.18575032 in the Zenodo[42].

## Code availability

Codes used in the current study are available at 10.5281/zenodo.18575032 in the Zenodo[42].

## Acknowledgements

We acknowledge financial support from the National Research Foundation of Korea (NRF) through the Basic Research Laboratory (No. RS-2024-00397664), Innovation Research Center (No. RS-2024-00413957), Young Researcher Program (No. RS-2025-00552989), Pilot and Feasibility Grants (No. RS-2025-19912971), and Midcareer Researcher Program (No. RS-2023-00274348), all funded by the Korean government. This work was supported by Creative-Pioneering Researchers Program and the BK21 FOUR program of the Education and Research Program for Future ICT Pioneers in 2026, through Seoul National University. We also acknowledge an administrative support from SOFT foundry institute.

## Author contributions

S.Yu, X.P., and N.P. conceived the idea of generalizing HU and SHU to non-Hermitian wave physics and supervised the project. S.Youn, G.L., and I.L. carried out the theoretical work and performed the numerical analyses. K.P. contributed to discussions on the practical implementation of the proposed design. All authors contributed to discussions of the results and to the writing of the manuscript.

## Competing interests

The authors have no conflicts of interest to declare.

# Supplementary Information for "Non-Hermitian stealthy hyperuniformity"

Gitae Lee[1‡], Seungmok Youn[1‡], Ikbeom Lee[1‡], Kunwoo Park[1], Duhwan Hwang[1], Xianji Piao[2§], Namkyoo Park[3†], and Sunkyu Yu[1*]

[1]Intelligent Wave Systems Laboratory, Department of Electrical and Computer Engineering, Seoul National University, Seoul 08826, Korea

[2]Wave Engineering Laboratory, School of Electrical and Computer Engineering, University of Seoul, Seoul 02504, Korea

[3]Photonic Systems Laboratory, Department of Electrical and Computer Engineering, Seoul National University, Seoul 08826, Korea

E-mail address for correspondence: §piao@uos.ac.kr, †nkpark@snu.ac.kr, *sunkyu.yu@snu.ac.kr

Note S1. Structure factors for non-Hermitian materials

Note S2. Semi-analytical expressions of density fluctuations

Note S3. Loss function and orthogonality of $S_A$ and $S_C$

Note S4. Non-Hermitian perturbed lattices

Note S5. Full-wave simulations

Note S6. Capturing unidirectionality with two correlated windows

**Note S1. Structure factors for non-Hermitian materials**

In the weak scattering regime under a far-field measurement condition, the scattering intensity is proportional to the generalized structure factor $S(\mathbf{k})$ which is defined as[1,2]:

$$S(\mathbf{k}) = \frac{1}{V_\Omega}\left|\int V_a(\mathbf{r})e^{-i\mathbf{k}\cdot\mathbf{r}}d\mathbf{r}\right|^2. \tag{S1}$$

Expanding Eq. (S1) with $V_a(\mathbf{r}) = V_r(\mathbf{r}) + iV_i(\mathbf{r})$ yields:

$$S(\mathbf{k}) = \frac{1}{V_\Omega}\left(|V_r(\mathbf{k})|^2 + |V_i(\mathbf{k})|^2\right) + \frac{2}{V_\Omega}\text{Im}\left[V_r(\mathbf{k})V_i^*(\mathbf{k})\right], \tag{S2}$$

where $V_r(\mathbf{k})$ and $V_i(\mathbf{k})$ denote the reciprocal-space potentials of $V_r(\mathbf{r})$ and $V_i(\mathbf{r})$, respectively. To analyse finite-size samples in view of the thermodynamic limit, ergodicity requires calculating statistical average of the samples[3], leading to Eq. (1) in the main text.

## Note S2. Semi-analytical expressions of density fluctuations

In this note, we derive semi-analytical expressions for auto- and cross-density fluctuations in statistically homogeneous non-Hermitian materials, providing a comprehensive characterization of their dependence on the window scale $R$.

### Cross-density fluctuation of the real and imaginary potential ($\sigma_C^2$)

Based on the real and imaginary potential perturbations in Eq. (4) of Methods, first consider the estimation of the cross-density fluctuation, $\sigma_C^2(R) = \langle N_r N_i \rangle - \langle N_r \rangle \langle N_i \rangle$, which can be written as:

$$\sigma_C^2(R) = f_r f_i \left\langle \sum_p s_p w^2(\mathbf{r}_p - \mathbf{x}_0; R) \right\rangle + f_r f_i \left\langle \sum_{p \neq q} s_p w(\mathbf{r}_p - \mathbf{x}_0; R) w(\mathbf{r}_q - \mathbf{x}_0; R) \right\rangle \quad (S3)$$
$$- f_r f_i \left\langle \sum_p w(\mathbf{r}_p - \mathbf{x}_0; R) \right\rangle \left\langle \sum_p s_p w(\mathbf{r}_p - \mathbf{x}_0; R) \right\rangle.$$

To evaluate Eq. (S3), we introduce an augmented probability space $(\mathbf{r}, s)$, where $s = \pm 1$ labels the sign of the imaginary potential. The first moments appearing in Eq. (S3) are given by

$$\left\langle \sum_p w(\mathbf{r}_p - \mathbf{x}_0; R) \right\rangle = \sum_{s \in \{-1,1\}} \int \rho(\mathbf{r}, s) w(\mathbf{r} - \mathbf{x}_0; R) d\mathbf{r}$$
$$= (\rho_{-1} + \rho_1) A(R) = \rho A(R), \quad (S4)$$

$$\left\langle \sum_p s_p w(\mathbf{r}_p - \mathbf{x}_0; R) \right\rangle = \sum_{s \in \{-1,1\}} \int \rho(\mathbf{r}, s) s \cdot w(\mathbf{r} - \mathbf{x}_0; R) d\mathbf{r}$$
$$= (\rho_1 - \rho_{-1}) A(R) = \langle s_p \rangle \rho A(R), \quad (S5)$$

which simplify the first and third terms in Eq. (S3) with $w(\mathbf{r} - \mathbf{x}_0; R) = w^2(\mathbf{r} - \mathbf{x}_0; R)$ and the constant number density $\rho_s = \rho(\mathbf{r}, s)$ under the statistical homogeneity.

The second term of Eq. (S3) involves the two-particle correlation. From the following Campbell's theorem on the second momentum:

$$\left\langle \sum_{p \neq q} f(\mathbf{r}_p, \mathbf{r}_q) \right\rangle = \sum_{s,s'} \int f(\mathbf{x}, \mathbf{y}) \rho^{(2)}(\mathbf{x}, s, \mathbf{y}, s') d\mathbf{x} d\mathbf{y}, \tag{S6}$$

where $\rho^{(2)}(\mathbf{x},s,\mathbf{y},s')$ denotes the second-order product density in the augmented space—that is, the product of the density values at **x** and **y** with corresponding $s$ and $s'$—we obtain

$$\left\langle \sum_{p \neq q} s_p w(\mathbf{r}_p - \mathbf{x}_0; R) w(\mathbf{r}_q - \mathbf{x}_0; R) \right\rangle$$
$$= \sum_{s \in \{-1,1\}} s \sum_{s' \in \{-1,1\}} \int w(\mathbf{x}; R) w(\mathbf{y}; R) \rho^{(2)}(\mathbf{x}, s, \mathbf{y}, s') d\mathbf{x} d\mathbf{y}. \tag{S7}$$

Due to the statistical homogeneity, this quantity can be expressed as

$$\int w(\mathbf{x};R) w(\mathbf{y};R) \rho^{(2)}(\mathbf{x},s,\mathbf{y},s') d\mathbf{x} d\mathbf{y}$$
$$= \rho^2 \int w(\mathbf{x};R) w(\mathbf{y};R) g_2(\mathbf{x}-\mathbf{y},s,s') d\mathbf{x} d\mathbf{y} \tag{S8}$$
$$= \rho^2 \int w(\mathbf{u};R) w(\mathbf{u}-\mathbf{r};R) g_2(\mathbf{r},s,s') d\mathbf{u} d\mathbf{r},$$

where $g_2(\mathbf{r},s,s')$ is the augmented pair correlation function. For fixed $s$ and $s'$, the spatial integral can be written in terms of the scaled intersection area,

$$\left\langle \sum_{p \neq q} s_p w(\mathbf{r}_p - \mathbf{x}_0; R) w(\mathbf{r}_q - \mathbf{x}_0; R) \right\rangle$$
$$= \rho^2 A(R) \int \alpha(\mathbf{r};R) \sum_{s,s'} s g_2(\mathbf{r},s,s') d\mathbf{r} \tag{S9}$$
$$= \rho^2 A(R) \int \alpha(\mathbf{r};R) \{g_2(\mathbf{r},1) - g_2(\mathbf{r},-1)\} d\mathbf{r},$$

where the dependences on the window shape and size are fully captured by the scaled intersection area $\alpha(\mathbf{r}; R)$, as

$$\alpha(\mathbf{r};R) = \frac{A_2^{\text{int}}(\mathbf{r};R)}{A(R)}, \quad A_2^{\text{int}}(\mathbf{r};R) \triangleq \int w(\mathbf{u};R) w(\mathbf{u}-\mathbf{r};R) d\mathbf{u}. \tag{S10}$$

We further define $g_2(\mathbf{r},s) \triangleq g_2(\mathbf{r},s,1) + g_2(\mathbf{r},s,-1)$, which encodes correlations as a function of relative position and the sign of the imaginary part at the destination. The cross-density fluctuation then becomes:

$$\frac{\sigma_C^2(R)}{\rho A(R)} = f_r f_i \langle s_p \rangle \{1 - \rho A(R)\} + f_r f_i \rho \int \alpha(\mathbf{r}; R) \{g_2(\mathbf{r},1) - g_2(\mathbf{r},-1)\} d\mathbf{r}. \qquad (S11)$$

**Auto-density fluctuation of the real potential ($\sigma_{AR}^2$)**

The auto-density fluctuation of real potential, $\sigma_{AR}^2$, is defined by $\langle N_r^2 \rangle - \langle N_r \rangle^2$, leading to:

$$\begin{aligned}\sigma_{AR}^2(R) &= f_r^2 \left\langle \sum_p w^2(\mathbf{r}_p - \mathbf{x}_0; R) \right\rangle + f_r^2 \left\langle \sum_{p \neq q} w(\mathbf{r}_p - \mathbf{x}_0; R) w(\mathbf{r}_q - \mathbf{x}_0; R) \right\rangle \\ &- f_r^2 \left\langle \sum_p w(\mathbf{r}_p - \mathbf{x}_0; R) \right\rangle \left\langle \sum_p w(\mathbf{r}_p - \mathbf{x}_0; R) \right\rangle.\end{aligned} \qquad (S12)$$

Because the imaginary components are irrelevant and all real components are identical, the dependence on $s$ and $s'$ can be eliminated by defining $g_2(\mathbf{r}) \triangleq \Sigma_{s,s'} g_2(\mathbf{r},s,s')$. Using the following equation:

$$\left\langle \sum_{p \neq q} w(\mathbf{r}_p - \mathbf{x}_0; R) w(\mathbf{r}_q - \mathbf{x}_0; R) \right\rangle = \rho^2 A(R) \int g_2(\mathbf{r}) \alpha(\mathbf{r}; R) d\mathbf{r}, \qquad (S13)$$

which is the direct consequence of the Campbell's theorem of the second momentum, Eq. (S12) reduces to

$$\frac{\sigma_{AR}^2(R)}{\rho A(R)} = f_r^2 \left[ 1 + \rho \int (g_2(\mathbf{r}) - 1) \alpha(\mathbf{r}; R) d\mathbf{r} \right]. \qquad (S14)$$

**Auto-density fluctuation of the imaginary potential ($\sigma_{AI}^2$)**

We consider the auto-density fluctuation of the imaginary potential $\sigma_{AI}^2$, defined as $\langle N_i^2 \rangle - \langle N_i \rangle^2$, which can be expanded as:

$$\sigma_{AI}^2(R) = f_i^2 \left\langle \sum_p s_p^2 w^2(\mathbf{r}_p - \mathbf{x}_0; R) \right\rangle + f_i^2 \left\langle \sum_{p \neq q} s_p s_q w(\mathbf{r}_p - \mathbf{x}_0; R) w(\mathbf{r}_q - \mathbf{x}_0; R) \right\rangle$$

$$- f_i^2 \left\langle \sum_p s_p w(\mathbf{r}_p - \mathbf{x}_0; R) \right\rangle \left\langle \sum_p s_p w(\mathbf{r}_p - \mathbf{x}_0; R) \right\rangle \tag{S15}$$

$$= f_i^2 \rho A(R) + f_i^2 \left\langle \sum_{p \neq q} s_p s_q w(\mathbf{r}_p - \mathbf{x}_0; R) w(\mathbf{r}_q - \mathbf{x}_0; R) \right\rangle - f_i^2 \langle s_p \rangle^2 \{\rho A(R)\}^2.$$

The second term in Eq. (S15) is calculated using $\rho^{(2)}$ as follows:

$$\left\langle \sum_{p \neq q} s_p s_q w(\mathbf{r}_p - \mathbf{x}_0; R) w(\mathbf{r}_q - \mathbf{x}_0; R) \right\rangle = \int w(\mathbf{x}; R) w(\mathbf{y}; R) \sum_{s,s'} ss' \rho^{(2)}(\mathbf{x} - \mathbf{y}, s, s') d\mathbf{x} d\mathbf{y}. \tag{S16}$$

We define the correlation function $c(\mathbf{r})$ as

$$c(\mathbf{r}) \equiv \sum_{s,s'} ss' g_2(\mathbf{r}, s, s') = \frac{1}{\rho^2} \sum_{s,s'} ss' \rho^{(2)}(\mathbf{r}, s, s'), \tag{S17}$$

which captures correlations in both spatial positions and imaginary components. Using this definition, Eq. (S16) can be rewritten in terms of the scaled intersection area,

$$\left\langle \sum_{p \neq q} s_p s_q w(\mathbf{r}_p - \mathbf{x}_0; R) w(\mathbf{r}_q - \mathbf{x}_0; R) \right\rangle$$

$$= \rho^2 \int w(\mathbf{u}; R) w(\mathbf{u} - \mathbf{r}; R) c(\mathbf{r}) d\mathbf{u} d\mathbf{r} = \rho^2 A(R) \int \alpha(\mathbf{r}; R) c(\mathbf{r}) d\mathbf{r}, \tag{S18}$$

which leads to the final expression, as follows:

$$\frac{\sigma_{AI}^2(R)}{\rho A(R)} = f_i^2 \left[ 1 + \rho \int \left[ c(\mathbf{r}) - \langle s_p \rangle^2 \right] \alpha(\mathbf{r}; R) d\mathbf{r} \right]. \tag{S19}$$

**Interpretation of density fluctuations with analytical forms**

Figure S1 compares the semi-analytical calculation of density fluctuations, derived from Eq. (S11), (S14), and (S19), with direct numerical measurements obtained in Fig. 2 of the main text. The comparison is shown for three representative classes of point patterns: a PT-symmetric crystal (Fig. S1a), a disordered non-Hermitian HU material (Fig. S1b), and a non-Hermitian Poisson random material (Fig. S1c). In all cases, the semi-analytical results quantitatively reproduce the numerical measurements over the entire range of window sizes, demonstrating the equivalence of the two estimators and validating the semi-analytical framework developed in this note.

Our semi-analytical framework reveals further design freedom for tailoring the microstructures of non-Hermitian materials by leveraging their cross-density fluctuations. As shown in Eq. (S11), the vanishing of $\sigma_C^2$ originates from the balanced numbers of gain and loss particles with $\langle s_p \rangle = 0$ and their identical statistical distributions with $g_2(\mathbf{r},1) = g_2(\mathbf{r},-1)$. From this observation, we present an example of non-Hermitian HU materials that exhibits nonzero cross-density fluctuations at finite $R$ by employing distinct statistical distributions for the gain and loss scatterers (Fig. S1d,e).

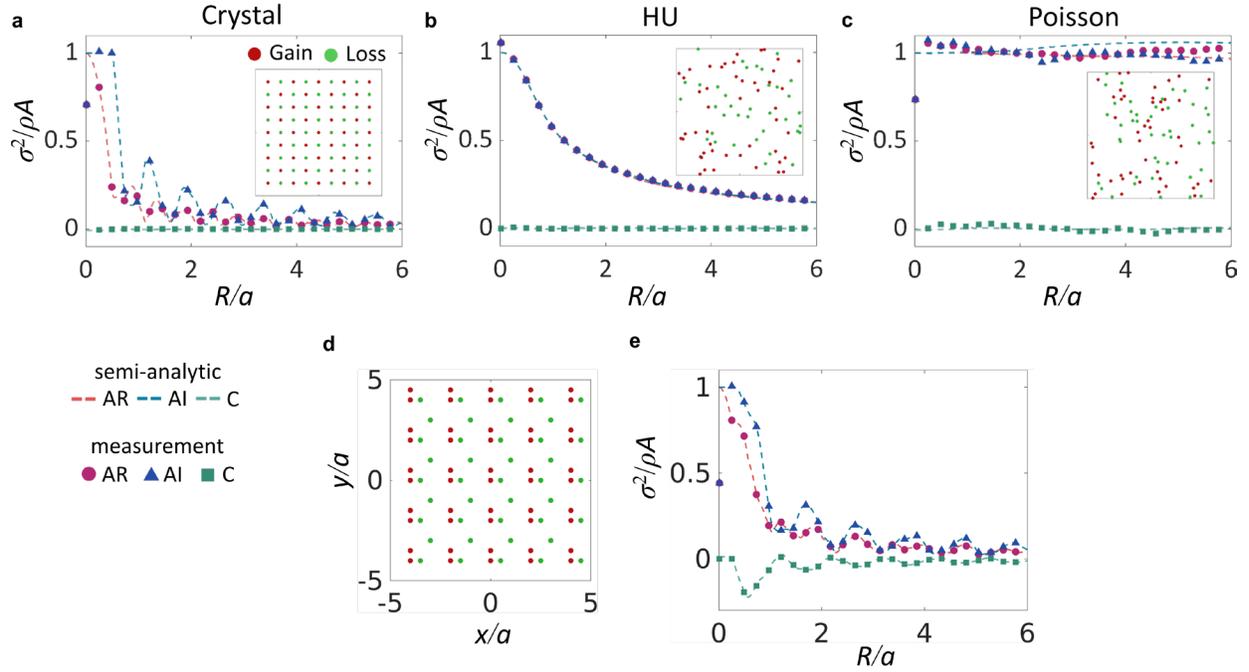

**Fig. S1. Semi-analytical density fluctuations in non-Hermitian point patterns. a-c,** Comparison between semi-analytical predictions (dashed lines) and numerical measurements (markers) for a square PT-symmetric lattice (**a**), a disordered non-Hermitian HU (**b**), and a non-Hermitian Poisson-random (**c**) material, corresponding to the structures shown in Fig. 2 of the main text. Insets show representative real-space configurations. The excellent agreement across all three classes confirms the validity of the semi-analytical estimator. (**d**) Non-Hermitian HU material with nonzero cross-density at finite $R$ (**e**).

**Note S3. Loss function and orthogonality of $S_A$ and $S_C$**

To generate point patterns with a designed structure factor, we minimize a weighted least-squares objective

$$\mathcal{L} = \sum_{\mathbf{k} \in B} \omega(|\mathbf{k}|)(S(\mathbf{k}) - S_0(\mathbf{k}))^2, \tag{S20}$$

using radial weights $\omega(|\mathbf{k}|) = 1/|\mathbf{k}|^2$, design region $B = \{\mathbf{k} \mid \mathbf{k} = 2\pi(n_x, n_y)/L,\ n_x, n_y \in \mathbb{Z},\ 0 < |\mathbf{k}| \leq K\}$, and a target structure factor $S_0 = S_{0,A} + S_{0,C}$. The corresponding weighted inner product is defined as:

$$\langle f, g \rangle_\omega \equiv \sum_{\mathbf{k} \in B} \omega(|\mathbf{k}|) f(\mathbf{k}) g(\mathbf{k}). \tag{S21}$$

By pairing $(\mathbf{k}, -\mathbf{k})$ and using the relationships, $S_A(-\mathbf{k}) = S_A(\mathbf{k})$ and $S_C(-\mathbf{k}) = -S_C(\mathbf{k})$, we can utilize $\langle S_A, S_C \rangle_\omega = 0$. Consequently, the loss function separates additively as

$$\mathcal{L} = \|S_A - S_{0,A}\|^2 + \|S_C - S_{0,C}\|^2, \tag{S22}$$

where $\|f\|$ denotes the weighted norm defined by $\langle f, f \rangle_\omega$. This orthogonality guarantees independent convergence of $S_A$ and $S_C$ toward their respective target ansatzes, $S_{0,A}$ and $S_{0,C}$, respectively.

## Note S4. Non-Hermitian perturbed lattices

Materials with unidirectional scattering properties at long wavelengths can also be realized by perturbing non-Hermitian lattices (Fig. S2). Depending on the propagation direction, the resulting materials exhibit an angle-dependent unsuppressed region of the structure factor near the origin of reciprocal space. These materials can be regarded as the defected-crystalline counterparts of non-Hermitian disordered SHU systems.

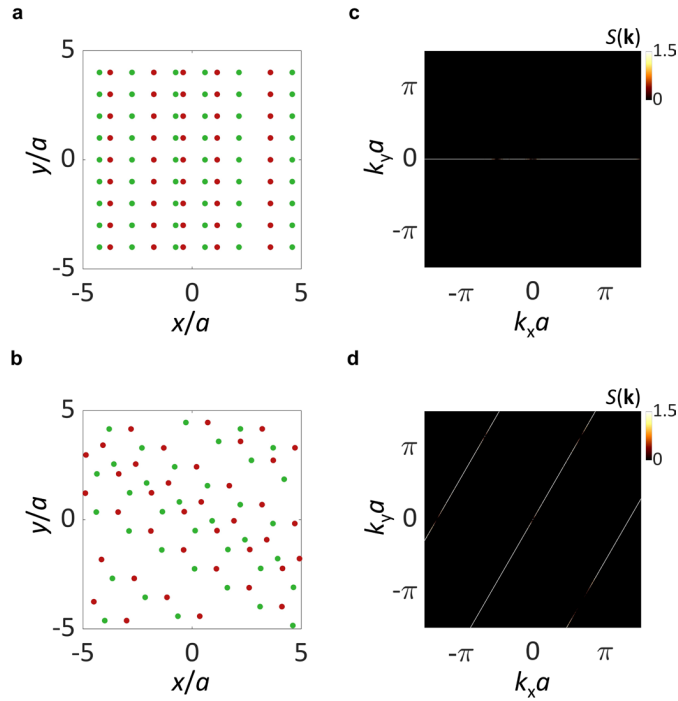

**Fig. S2. Non-Hermitian perturbed lattices. a,b,** Realizations of perturbed non-Hermitian lattices, where perturbations are applied on the square lattice (**a**) and a hexagonal lattice (**b**), which correspond to the crystalline structures shown in Fig. 3f and Fig. 3g of the main text, respectively. **c-d,** Corresponding structure factors of perturbed square (**c**) and hexagonal (**d**) lattices.

**Note S5. Full-wave simulations**

We validate our analysis using full-wave finite-difference time-domain (FDTD) simulations performed with Tidy3D[4]. The realization generated in a dimensionless square domain $[0, L]^2$ is mapped to physical coordinates through a scale factor $\xi = K\lambda_0/(2\pi)$, where $\lambda_0 = 0.5$ $\mu$m denotes the free-space wavelength. This scaling ensures that the accessible momentum-transfer grid, expressed in units of $k_0=2\pi/\lambda_0$, spans $\mathbf{k}/k_0 \in [-2K, 2K]^2$. The physical size of the scattering region becomes $L_{SC} = L\xi$.

Each scatterer is modelled as a cylindrical rod of radius $r_{rod} = \lambda_0/45$, positioned at the optimized locations $\mathbf{r}_p$. The scatterers possess a complex refractive index $n_p = 1 + \Delta n + i\kappa_p$, with $\Delta n = 0.05$ and $\kappa_p \in \{0.05, -0.05\}$. The background medium is air ($n = 1$). We employ a uniform Cartesian grid with a spatial resolution of $r_{rod}/7$. The structure is excited using a plane-wave total-field/scattered-field (TFSF) source at centre frequency $f_0 = c/\lambda_0$, with a fractional bandwidth of $\Delta f/f_0 = 1/20$. The polarization is set to the transverse-electric mode.

The azimuthal angle of incidence $\theta_{inc}$ is swept from 0° to 357° in steps of 3°. Four rectangular field monitors (top, bottom, left, and right) are placed outside the TFSF boundary at the distance of $0.2\lambda_0$. The recorded near fields are projected onto a far-field angular ring using Tidy3D's angle projector[4], sampling observation angles $\theta \in [0, 2\pi]$ with a resolution of 3°. Each simulation runs for a total time of $5L_{SC}/c$, which is sufficient to ensure convergence.

The projected far-field intensity measured at angle $\theta$ for a plane-wave incidence from direction $\theta_{inc}$, denoted as $I(\theta; \theta_{inc})$, is normalized by the scattering intensity $I_0(\theta)$—defined as the sum of the scattering intensities of individual gain and loss—to obtain the structure factor:

$$S(\mathbf{k}) \sim \frac{I(\theta; \theta_{inc})}{I_0(\theta; \theta_{inc})}. \tag{S23}$$

Finally, we decompose the simulated structure factor into its auto- and cross- components: $S_A(\mathbf{k})$ = $[S(\mathbf{k}) + S(-\mathbf{k})]/2$, $S_C(\mathbf{k}) = [S(\mathbf{k}) - S(-\mathbf{k})]/2$. For an inversely designed sample with $\alpha = 0.2$, consisting of 2000 gain and 2000 loss scatterers, we truncate a circular region in real space and obtain $S(\mathbf{k})$ (Fig. S3a,b), $S_A(\mathbf{k})$ (Fig. S3c,d), and $S_C(\mathbf{k})$ (Fig. S3e,f). Structure factors of $m = 2$ (Fig. S3a,c,e) and $m = 3$ (Fig. S3b,d,f) show clear 2- and 3-fold rotational symmetry of suppression region, respectively, as demonstrated in the main text.

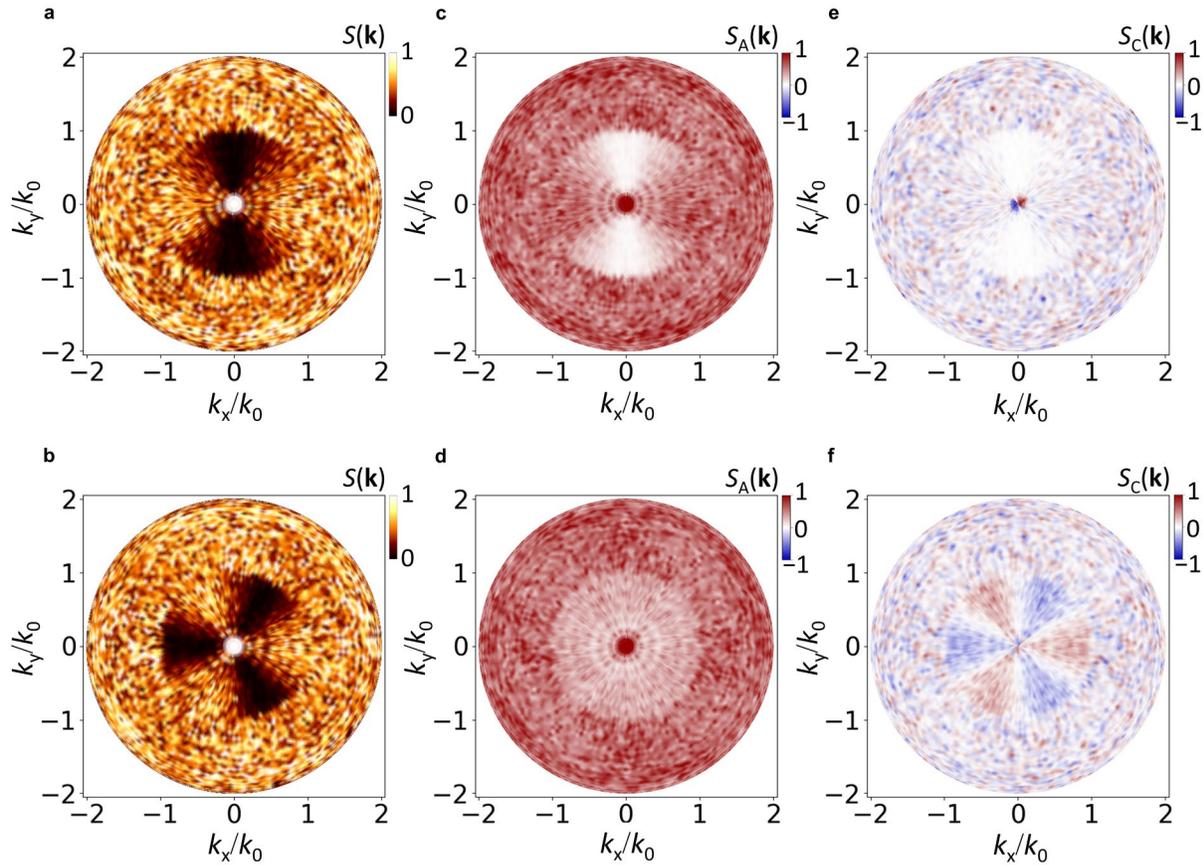

**Fig. S3. Structure factors obtained with full-wave analysis.** Total structure factors $S(\mathbf{k})$ for $m = 2$ (**a**) and $m = 3$ (**b**). Corresponding $S_A(\mathbf{k})$ (**c,d**) and $S_C(\mathbf{k})$ (**e,f**). To ensure statistical robustness, 5 independent sample realizations are averaged.

**Note S6. Capturing unidirectionality with two correlated windows**

The odd component of the structure factor, $S_C(\mathbf{k})$, which underlies unidirectional scattering, can be expressed as the Fourier transform of the real-imaginary correlation function $\rho_{ri,2}(\mathbf{r})$:

$$\rho_{ri,2}(\mathbf{r}) \triangleq \int V_r(\mathbf{r}')V_i(\mathbf{r}'+\mathbf{r})d\mathbf{r}',$$
$$S_C(\mathbf{k}) \propto \int \left(\rho_{ri,2}(\mathbf{r}) - \rho_{ri,2}(-\mathbf{r})\right)e^{i\mathbf{k}\cdot\mathbf{r}}d\mathbf{r} = 2\int \mathcal{O}\left(\rho_{ri,2}(\mathbf{r})\right)e^{i\mathbf{k}\cdot\mathbf{r}}d\mathbf{r},$$
(S24)

where $\mathcal{O}\{\cdot\}$ denotes the odd part of a function. Because $S_C(\mathbf{k})$ is directly related to the odd component of $\rho_{ri,2}(\mathbf{r})$, conventional sampling schemes employing a single window fail to capture it due to the even-functional property of the associated kernel:

$$\langle N_r(\mathbf{r};\mathbf{R})N_i(\mathbf{r};\mathbf{R})\rangle = A(\mathbf{R})\int d\mathbf{r}\,\rho_{ri,2}(\mathbf{r})\alpha(\mathbf{r};\mathbf{R}),$$
(S25)

where $\mathbf{R}$ specifies the orientation and size of a window of given shape, $N_r(\mathbf{r};\mathbf{R})$ and $N_i(\mathbf{r};\mathbf{R})$ denote the net real and imaginary potential perturbations within a window placed at $\mathbf{r}$, respectively, and $\alpha(\mathbf{r};\mathbf{R})$ represents the scaled intersection volume of windows with volume $A(\mathbf{R})$. Because $\alpha(\mathbf{r};\mathbf{R}) = \alpha(-\mathbf{r};\mathbf{R})$, the odd component of $\rho_{ri,2}(\mathbf{r})$ cannot be detected. To overcome this limitation, we introduce two distinct but spatially correlated windows in each trial, which generally breaks the even-functional property of $\alpha(\mathbf{r};\mathbf{R})$ and allows odd contributions:

$$\langle N_{1,r}(\mathbf{r};\mathbf{R})N_{2,i}(\mathbf{r};\mathbf{R})\rangle = A(\mathbf{R})\int d\mathbf{r}\,\rho_{ri,2}(\mathbf{r})\alpha_{12}(\mathbf{r};\mathbf{R}),$$
$$\alpha_{12}(\mathbf{r};\mathbf{R}) = \frac{1}{A(\mathbf{R})}\int d\mathbf{r}'\,w_1(\mathbf{r}';\mathbf{R})w_2(\mathbf{r}'+\mathbf{r};\mathbf{R}).$$
(S26)

In general, the resulting cross-kernel, $\alpha_{12}(\mathbf{r};\mathbf{R})$, is not an even function, even for circular windows. Specifically, we employ two circular windows of the identical radius $a$, the characteristic length, whose centres are separated by relative displacement $\mathbf{R}_c = r(\hat{\mathbf{x}}\cos\varphi + \hat{\mathbf{y}}\sin\varphi)$. The first window only measures the net real potential perturbation within it, denoted as $N_{1,r}$, while the second window measures only the net imaginary potential perturbation, denoted as $N_{2,i}$. The odd-part of real-imaginary correlation function is then captured by trial measurements while varying the direction and norm of $\mathbf{R}_c$.